\documentclass[pra,twocolumn,showpacs]{revtex4}
\usepackage{amsmath,amssymb}
\usepackage[dvips]{graphicx}
\usepackage[FIGTOPCAP]{subfigure}
\newcommand \beq{\begin{eqnarray}}
\newcommand \eeq{\end{eqnarray}}

\begin{document}
\title{Spin-Depairing Transition of 
Attractive Fermi Gases on a Ring \\  Driven by
 Synthetic Gauge Fields}
\author{Shun Uchino and Norio Kawakami}
\affiliation{Department of Physics, Kyoto University, Kyoto 606-8502, Japan}
\begin{abstract}
Motivated by the recent experimental realization of synthetic gauge fields
in ultracold atoms, we investigate one-dimensional attractive 
Fermi gases
with a time-dependent gauge flux on the spin sector.
By combining  the methods of the Bethe ansatz  
with complex twists and Landau-Dykhne, it is shown that 
a spin-depairing transition occurs,
which may represent a nonequilibrium transition from
fermionic superfluids to normal states with spin currents
caused by a many-body quantum tunneling.
For the case of the Hubbard ring at half filling,
our finding forms a dual concept with the dielectric breakdown
of the Mott insulator discussed in Phys. Rev. B \textbf{81}, 033103
 (2010).
We analyze cases of arbitrary filling and continuum model, and 
show how filling affects the transition probability.

\end{abstract}
\pacs{03.65.Xp, 05.60.Gg, 67.85.-d}
\maketitle
\section{Introduction}
Owing to high controllability and 
flexibility of experiments,
ultracold atoms have offered a testing ground 
to simulate a strongly correlated system in one dimension both under 
equilibrium and nonequilibrium conditions \cite{bloch,cazalilla}.
Recent experimental efforts make it possible to study 
the system on a ring, which has been  
realized in continuum space \cite{gupta,ryu} and 
in optical lattice \cite{henderson}.
As far as two-component Fermi gases on a ring in equilibrium are concerned,
one may obtain exact solutions 
by the Bethe ansatz method regardless of continuum \cite{yang} or 
lattice \cite{lieb}, and many of the properties of the systems
can be examined in combination with the bosonization approach. 
However, as for what happens in the Fermi gases in nonequilibrium,
a unified approach has yet to be established
and our current knowledge is far from a comprehensive level. 

Meanwhile, in recent years studies of synthetic
gauge fields have attracted attention in cold-atom community \cite{dalibard}
and the NIST group has succeeded in producing synthetic
magnetic \cite{lin1} and electric fields \cite{lin3},
and spin-orbit coupling \cite{lin2} on a continuum space. 
In particular, the electric fields discussed in Ref. \cite{lin3}
have been realized with the creation of time-dependent gauge fields. 
In addition, very recently, Aidelsburger \textit{et al}. have succeeded in
creating synthetic gauge fields on an optical lattice
\cite{aidelsburger},
which may pave the way to areas of study in cold
atomic gases.  
In this paper, based on the situations described above,
we consider a two-component attractive Fermi gas on a ring
with a time-dependent gauge flux on the spin sector at absolute zero.
In the absence or presence of a time-independent gauge flux, 
it is well known that 
the ground state of the two-component attractive Fermi gas 
is filled with bound states of up-spin and
down-spin particles and
the spin excitation has a gap, which is attributed to the appearance of 
fermionic superfluidity \cite{essler2}.
Here, we demonstrate that in the presence of 
a time-dependent gauge flux $\Phi(t)=Ft$,
a spin-depairing transition induced by a many-body 
quantum tunneling occurs, which
may represent a breaking phenomenon of fermionic superfluidity
in nonequilibrium.
Our prediction relies on a recently proposed method \cite{oka,oka2} 
combining the Bethe ansatz method with
complex twists \cite{fukui,nakamura} 
and the Landau-Dykhne method \cite{landau,dykhne,davis} with which
a quantum tunneling probability between two states
can be evaluated.
In fact, for the Hubbard model, our finding builds a dual concept with the 
nonequilibrium  transition from 
Mott insulators to metals in the repulsive Hubbard ring at half filling
with a time-dependent charge flux discussed
in the context of electronic systems \cite{oka}.
In addition, we show that by using the formalism with dressed energies 
\cite{woynarovich,essler2},
the spin-depairing transition occurs 
not only at half filling but also other 
filling cases, and in continuum model. This is peculiar to the 
attractive Fermi gases and is in contrast with
the breakdown of the Mott insulator, which
occurs in lattice systems only at half filling.

This paper is organized as follows.  Section I\hspace{-.1em}I
discusses a Hamiltonian with synthetic gauge fields
and shows that effects of gauge fields can be incorporated 
into boundary conditions.
Section I\hspace{-.1em}I\hspace{-.1em}I
examines Bethe ansatz equations and spin gap
with complex spin twists.
We analytically show that the spin gap closes at a critical value. 
In Sec. I\hspace{-.1em}V, we relate the real spin twist to
the complex twist with the Landau-Dykhne method and
evaluate the transition probability. 
We also comment on the ground-state decay rate.
Section V summarizes the contributions and
examines the continuum model.

\section{Hamiltonian}
We start by considering a Hamiltonian of two component atoms 
influenced by synthetic gauge fields 
to yield the spin-orbit coupling. In the continuum space,
the kinetic term becomes the covariant form:
$-(\vec{\nabla}+i\vec{A})^2,$ 
where $\vec{A}=\sum_{a=1}^{3}\vec{A}_a(\vec{x},t)\tau^{a}$ 
represents the synthetic gauge field and $\tau^{a}$ is
the generator of SU(2).
We then wish to consider the situation that 
the particles are loaded into an optical lattice.
Let us assume that the lattice potential is so deep that 
the energy gap between first and second bands is much larger than
the other effects such as 
the thermal and mean-field interaction energies per particle.
If $V$ is the lattice potential and
$\omega$ is the Wannier function, which is localized at 
each lattice point,
the hopping amplitude becomes 
\beq
t'_{ij}&=&\int d\vec{x}\omega^{*}(\vec{x}-\vec{R}_i)[
-(\vec{\nabla}+i\vec{A})^2
+V(x)]\omega(\vec{x}-\vec{R}_j)\nonumber\\
&=&\int d\vec{x}\omega^{*}(\vec{x}-\vec{R}_i)S^{-1}(x,x_0)
[-\vec{\nabla}^2+V(x)]\nonumber\\
&&\times S(x,x_0)\omega(\vec{x}-\vec{R}_j)\cong S(i,j)t_{ij},
\label{eq:hopping-gf}
\eeq
where $t_{ij}=\int d\vec{x}\omega^{*}(\vec{x}-\vec{R}_i)[
-\vec{\nabla}^2+V(x)]\omega(\vec{x}-\vec{R}_j)$
is the hopping amplitude at $\vec{A}=0$,
\beq
S(i,j)
&=&[1+i\vec{A}(\vec{R}_i,t)\cdot\Delta\vec{y}
][1+i\vec{A}(\vec{R}_i+\Delta \vec{y},t)\cdot\Delta\vec{y}]\dots
\nonumber\\
&=&\lim_{N\to\infty}\prod_{n=0}^{N-1}[
1+i\vec{A}(\vec{R}_i+n\Delta \vec{y},t)\cdot\Delta\vec{y}]\nonumber\\
&\equiv&P \exp \Big[i\int_{\vec{R_i}}^{\vec{R_j}}
\vec{A}(\vec{y},t)\cdot d\vec{y}
\Big]
\label{eq:path-order}
\eeq
is the so-called Wilson line in gauge theory, 
and $P$ denotes the path-ordered product \cite{peskin}.
Here, to obtain the last expression of Eq. \eqref{eq:hopping-gf},
we used the assumption that the Wannier function is tightly localized at 
each lattice point.

Although the above discussion is independent of space dimensions,
geometry of systems, and statistics of particles,
we hereafter concentrate on a one-dimensional ring, and 
assume that the hopping is site-independent.
Then, the Hamiltonian of two-component Fermi gases 
with a contact attractive interaction
in units of the hopping strength is given by
\beq
H&=&-\sum_{i,\sigma,\sigma'}c^{\dagger}_{i\sigma}
S(i,i+1)_{\sigma\sigma'}c_{i+1\sigma'}
+\text{H.c.}\nonumber\\
&&-4|u|\sum_{i}n_{i\uparrow}n_{i\downarrow}.
\label{eq:so-hubbard-ring}
\eeq
We note that this Hamiltonian corresponds to that analyzed in Ref. 
\cite{fujimoto} except for points that 
the attractive case is concerned and there is no U(1) charge flux
in our treatment.
In the following, for simplicity we consider the case that 
one of $\vec{A}_a$, say, $\vec{A}_3$ has a nonzero value,
which is comparable to the case of SO(2) spin symmetry.
This simplified gauge configuration is also considered
in the continuum space in Refs. 
\cite{anderson,edge}.
Then, by considering the following transformation \cite{fujimoto}:
\beq
c_{i\sigma}=[S(i,i-1)S(i-1,i-2)\cdots
S(2,1)]_{\sigma\tilde{\sigma}}
c_{i\tilde{\sigma}},
\label{eq:unitary-trans}
\eeq
Eq. \eqref{eq:so-hubbard-ring} is rewritten as
follows:
\beq
H=-\sum_{i,\sigma}c^{\dagger}_{i\sigma}
c_{i+1\sigma}
+\text{H.c.}
-4|u|\sum_{i}n_{i\uparrow}n_{i\downarrow}
\label{eq:so-hubbard2}
\eeq
with
\beq
c_{L+1\sigma}=\exp[i\sigma\Phi L]c_{1\sigma},
\label{eq:twist}
\eeq
where $L$ is the number of sites, $\sigma=\pm 1$ (or
$\sigma$ represents $\uparrow$ or $\downarrow$), 
and $\Phi$ is the gauge flux per site. 
Equation \eqref{eq:twist} indicates that the effect of gauge fields
is incorporated into twisted boundary conditions.

\begin{figure}
\includegraphics[width=0.7\linewidth]{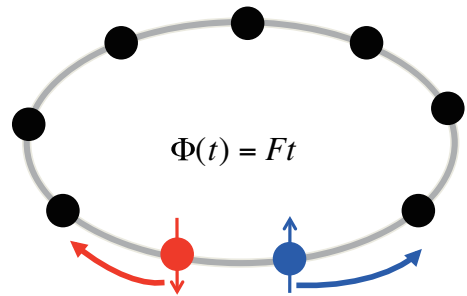}
\caption{(Color online)
Schematic illustration of spin currents induced 
by the time-dependent gauge flux $\Phi(t)=Ft$
on the spin sector. Each black dot without an arrow 
indicates a particle with spin-up and counterclockwise flow or
spin-down and clockwise flow.}
\label{current}
\end{figure}

Let us briefly comment on a role of $\Phi$.
As discussed in Refs. \cite{kohn,shastry,fujimoto}, 
in the repulsive Hubbard model,
$\Phi$ induces spin currents since $\Phi$ affects the spin sector and 
there is no spin gap.
In contrast, in the attractive Hubbard model without spin imbalance,
since there is a spin gap, spin currents do not flow
with a static $\Phi$, and therefore $\Phi$ does not play any role. 
In subsequent sections, however,
we show that the time-dependent gauge flux
$\Phi(t)=Ft$ induces spin currents through a many-body quantum tunneling
as shown in Fig. \ref{current}.

\section{Bethe Ansatz Equations with complex spin twists}
Let us consider the Hamiltonian \eqref{eq:so-hubbard2} 
with complex spin twists,
which is equivalent to the substitution $\Phi=i\Psi$ in 
Eq. \eqref{eq:twist} and is related to the repulsive Hubbard ring 
with complex charge twists.
As for complex charge twists, this model 
has been phenomenologically introduced
in Ref. \cite{fukui} to describe the dissipative tunneling into 
the environment and analyze the breakdown of the Mott insulator.
In fact, it has been shown that due to complex charge twists
the Mott gap closes, which evidences the breakdown of the Mott
insulator.
In this section, we discuss the spin-depairing transition 
due to complex spin twists.
The connection between the Hubbard ring with real twists 
and that with complex twists is shown in Sec. I\hspace{-.1em}V
based on the analysis in Ref. \cite{oka}.

Now we examine solutions of the Hubbard ring with complex spin twists.
With absence of the twists, our model reduces to the one-dimensional 
Hubbard model
with periodic boundary conditions, which can be solved exactly by
the Bethe ansatz method \cite{lieb}.
While in the presence of \eqref{eq:twist},
the boundary condition is altered from
periodic to twisted ones, we can still obtain exact solutions by
the same method \cite{shastry,fukui}, and
the corresponding Bethe ansatz equations 
are then for charge $k_j$ and spin $\lambda_{\alpha}$ rapidities,
\begin{gather}
e^{ik_jL}=e^{-\Psi L}\prod_{\beta=1}^{N_{\downarrow}}
\frac{\lambda_{\beta}-\sin k_j+i|u|}{\lambda_{\beta}-\sin k_j -i|u|},
\label{eq:bethe-twist-1}\\
\prod_{j=1}^{N}\frac{\lambda_{\alpha}-\sin k_j+i|u|}
{\lambda_{\alpha}-\sin k_j -i|u|}=-e^{2\Psi L}
\prod_{\beta=1}^{N_{\downarrow}}
\frac{\lambda_{\alpha}-\lambda_{\beta}+2i|u|}
{\lambda_{\alpha}-\lambda_{\beta}-2i|u|},
\label{eq:bethe-twist-2}
\end{gather}
where $N$ and $N_{\downarrow}$ are the numbers of
total particles and particles with down spin, respectively.
We note that the term $e^{2\Psi L}$ in Eq. \eqref{eq:bethe-twist-2}
is peculiar to the case of the spin twist.
The energy of the system is then given by
\beq
E=-2\sum_j\cos k_j.
\eeq

In this paper, we only consider the case of $N=2N_{\downarrow}$,
which implies that there is no spin imbalance.
When $\Psi=0$, the ground state is then filled with the
spin pairs, each of which forms the bound state of 
a particle with spin-up and a particle with
spin-down to be called
the $k-\lambda$ string \cite{essler2}:
\beq
\sin k_{\alpha}^{\pm}=\lambda_{\alpha}\pm i|u| +O(e^{-\eta L}),
\label{eq:string-1}
\eeq
where $\eta$ is assumed to be a positive number of the order of unity.
We see that Eq. \eqref{eq:string-1} is consistent with Eq. 
\eqref{eq:bethe-twist-2}.
When $\Psi\ne 0$, however, the $k-\lambda$ string solutions \eqref{eq:string-1}
no longer satisfy Eq. \eqref{eq:bethe-twist-2}.
At the same time, if $\Psi$ is less than a critical value, which 
is determined below,
it is expected that
$k-\lambda$ string solutions still exist in
the ground state. 
In fact, we find that the following $k-\lambda$ string trivially satisfies
Eq. \eqref{eq:bethe-twist-2}:
\beq
\sin k_{\alpha}^{\pm}=\lambda_{\alpha} \pm i|u|
+O(e^{-(\eta\mp\Psi)L}),
\label{eq:string-2}
\eeq
where $\eta\pm\Psi$ are assumed to be positive.
Namely, the modification only occurs in the order of $e^{-\eta L}$.
Taking into account the fact that
it is difficult to show the mathematical existence of
Eq. \eqref{eq:string-1},
the same difficulty also occurs in Eq. \eqref{eq:string-2}.
However, since we can check the consistency with the Hubbard ring with
complex charge twists at half filling where 
there is the SO(4) symmetry \cite{essler2}, 
 we believe that Eq. \eqref{eq:string-2} 
is correct at arbitrary filling.
On the other hand, Eq. \eqref{eq:bethe-twist-1}  becomes
\beq
e^{i(k_{\alpha}^{+}+k_{\alpha}^{-})L}=\prod_{\beta=1}^{N_{\downarrow}}
\frac{\lambda_{\alpha}-\lambda_{\beta}-2i|u|}
{\lambda_{\alpha}-\lambda_{\beta}+2i|u|}.
\label{eq:bethe-twist-4}
\eeq
By taking logarithm of Eq. \eqref{eq:bethe-twist-4},
we obtain
\beq
2L\text{Re}\left[\sin^{-1}(\lambda_{\alpha}+i|u|)\right]
=2\pi J_{\alpha}-\sum_{\beta=1}^{N_{\downarrow}}\theta\left(
\frac{\lambda_{\alpha}-\lambda_{\beta}}{2}\right),
\nonumber\\
\label{eq:gs}
\eeq
where $\displaystyle\theta(x)=-2\tan^{-1}\left(x/|u|\right)$ and
$J_{\alpha}$ is integer or half integer.
We note that since  
Eq. \eqref{eq:gs} does not depend on $\Psi$,
the configuration of spin rapidities at $\Psi\ne 0$
is equal to that at $\Psi=0$.
Considering that the string solutions 
require the tight condition between $k^{\pm}_{\alpha}$ and
$\lambda_{\alpha}$ and therefore
$\lambda_{\alpha}$ cannot be changed by $\Psi$,
this may be the natural result.

Let us next consider the spin excitation above the ground state.
As stated above, the spin excitation
has a gap to be attributed to the appearance of fermionic superfluidity.
In order to see what happens by complex twists,
let us consider a spin-triplet excitation, which is accomplished by
the manipulation breaking one of the $k-\lambda$ string pairs
and creating two charge rapidities $k_1$ and $k_2$ \cite{essler2}.
We find that the corresponding Bethe ansatz equations are given by
\begin{gather}
2L\text{Re}\left[\sin^{-1}(\lambda_{\alpha}+i|u|)\right]
=2\pi J_{\alpha}-\sum_{\beta=1}^{N_{\downarrow}-1}
\theta\left(\frac{\lambda_{\alpha}-\lambda_{\beta}}{2}\right)\nonumber\\
-\sum_{j=1}^{2}\theta(\lambda_{\alpha}-\sin k_j),
\label{eq:excitation-1}\\
k_jL=2\pi I_j+iL\Psi -\sum_{\beta=1}^{N_{\downarrow}-1}\theta(\sin k_j
-\lambda_{\beta}), \ \ \ (j=1 \ \text{or} \ 2)
\label{eq:excitation-2}
\end{gather}
where $J_{\alpha}$ and $I_{j}$ are integers or half integers.
Compared with the Bethe ansatz equations in the ground state,
it is clear that solutions of the above equations depend on $\Psi$ explicitly.
Since $\lambda_{\alpha}$ represent the $k-\lambda$ strings,
it is expected that the shift of the distribution of $\lambda_{\alpha}$ 
by $\Psi$ is tiny 
and  major contributions by $\Psi$ come from $k_j$. 
In fact,  
by solving the above equations for a finite system numerically, 
we confirmed them.
This is the same situation as in the repulsive Hubbard ring with complex
charge twists where the behaviors with respect to
$k_j$ and $\lambda_{\alpha}$ are interpreted as the so-called
charge-spin separation \cite{fukui}.
\begin{figure}
\subfigure[
 $L=2N_{\downarrow}=50$]{
\includegraphics[width=0.7\linewidth]{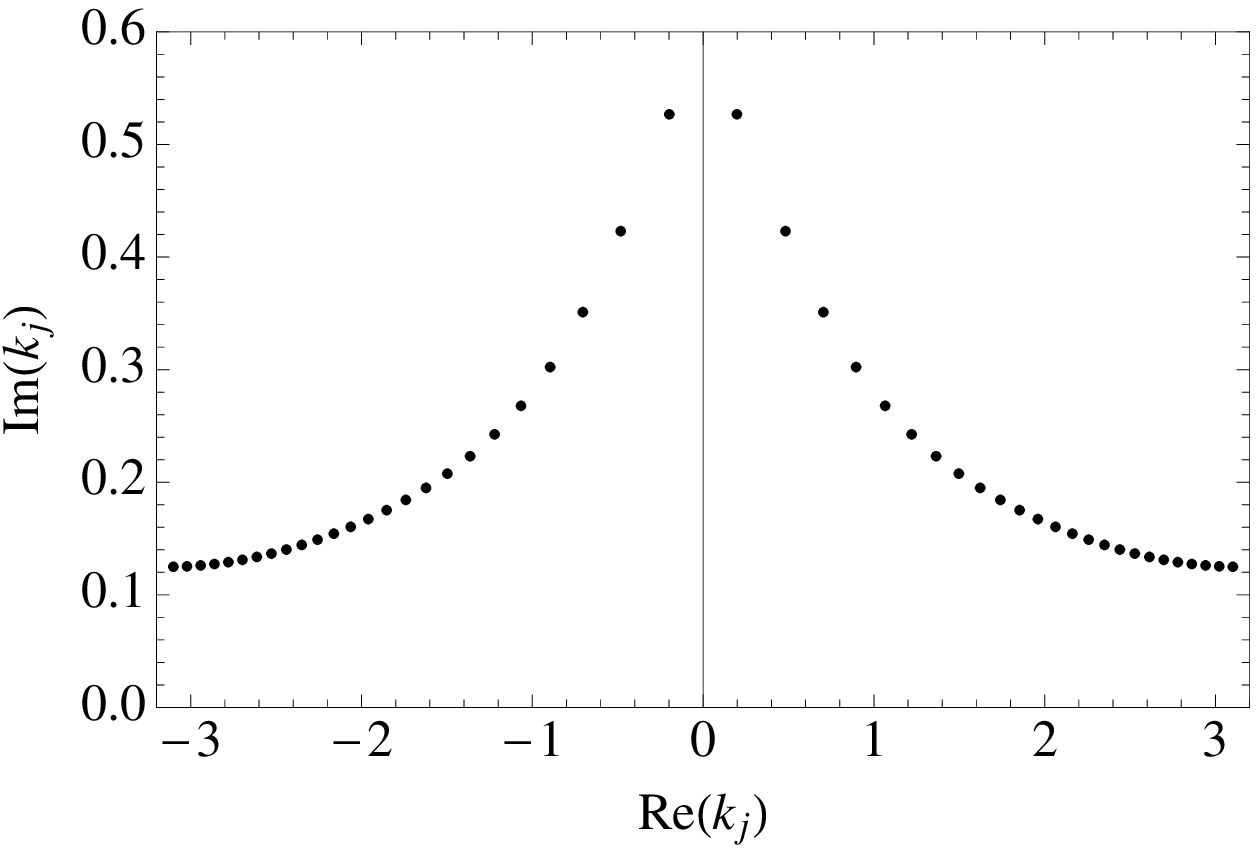}
}
\subfigure[$L=50, \ N_{\downarrow}=15$]{
 \includegraphics[width=0.7\linewidth]{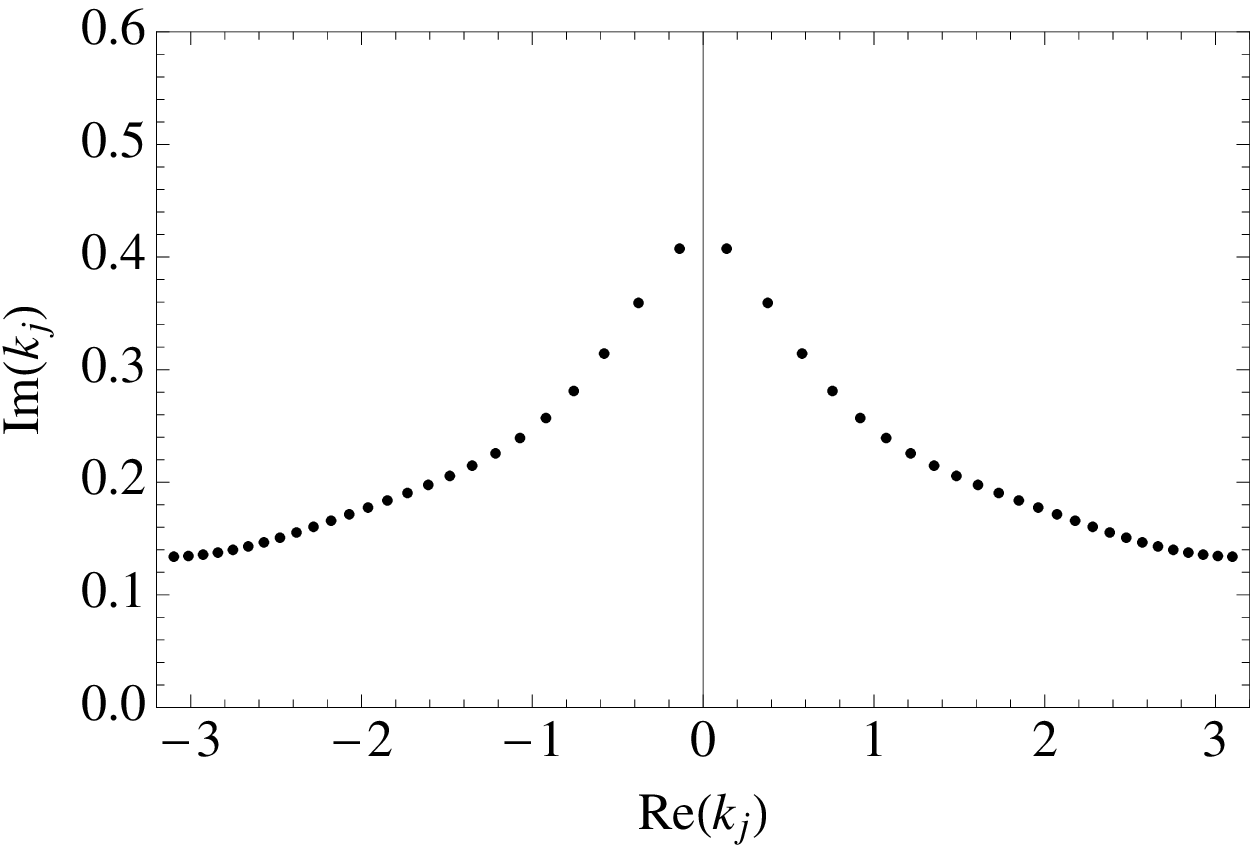}
}
\caption{Possible values of $k_j$ with $|u|=1.0$ and $\Psi=0.2$.
By comparing (a) with (b), we see that for fixed $\Psi$,
the maximum value of 
$\text{Im}(k_j)$ at half filling  is larger than that below
half filling.}
\label{spin-distribution}
\end{figure} 
As for $\text{Im}(k_j)$, regardless of filling
this takes a maximum value at 
$\text{Re}(k_j)=0$ as can be seen from Fig. \ref{spin-distribution}.
This contrasts to the repulsive Hubbard ring with complex charge
twists where $\text{Im}(k_j)$ takes a maximum value at 
$\text{Re}(k_j)=\pm\pi$.
This difference between repulsion and attraction is related to the
observation that at least for $\Psi=0$, the charge excitation 
of the repulsive Hubbard ring
corresponds  to the spin excitation of the attractive one by
the substitution $k\to \pi -k$ \cite{essler2}.
The result of Fig. \ref{spin-distribution} implies that this still holds
for $\Psi\ne 0$.

\subsection*{Analysis in the Bulk Limit}
\begin{figure}[t]
\subfigure[$|u|=0.25$]{
\includegraphics[width=0.75\linewidth]{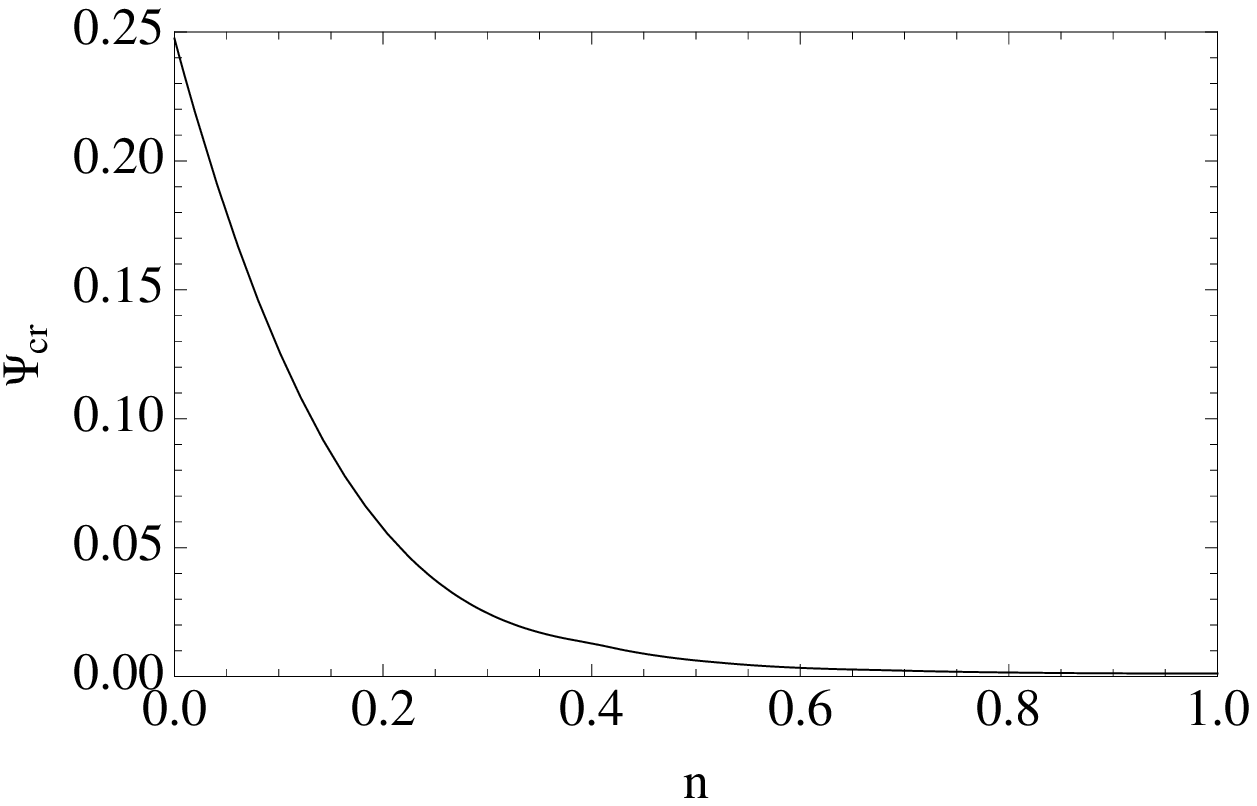}
}
\subfigure[$|u|=0.5$]{
\includegraphics[width=0.75\linewidth]{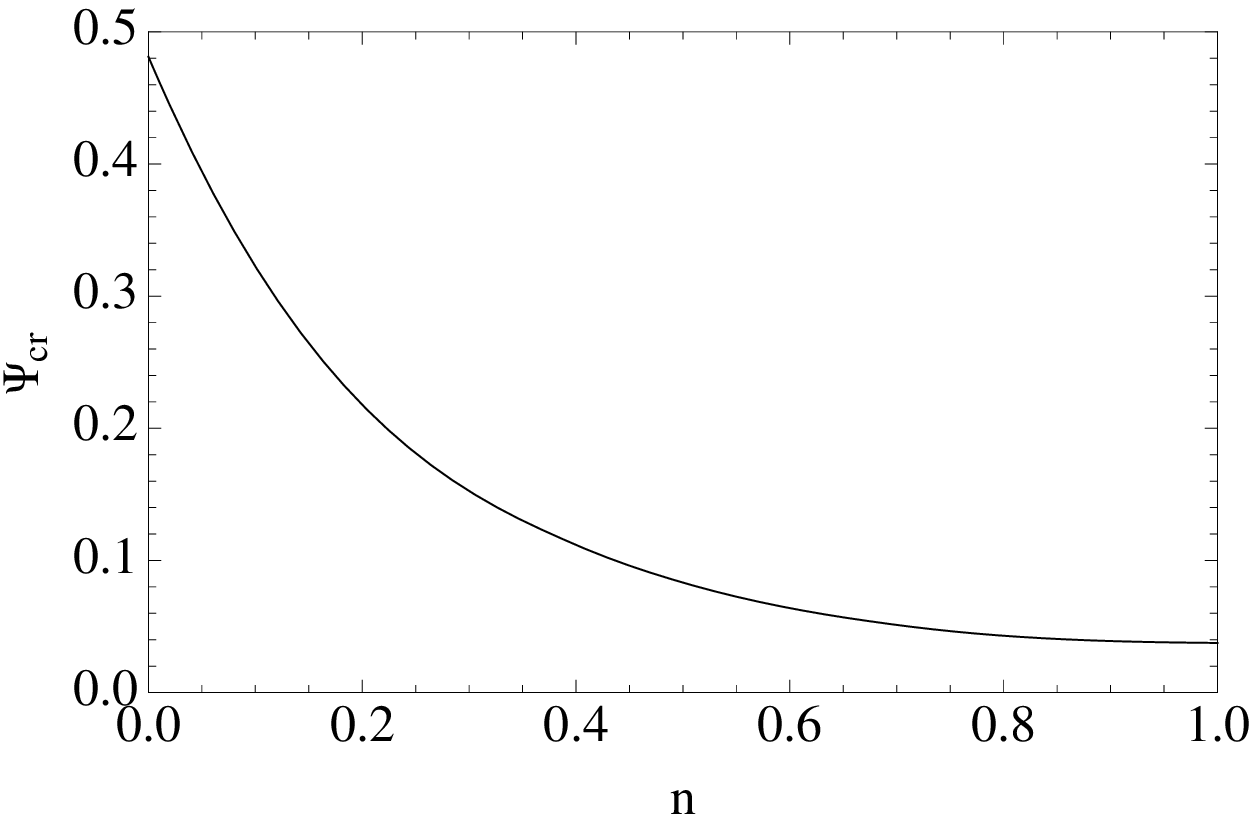}
}
\subfigure[$|u|=1.0$]{
 \includegraphics[width=0.75\linewidth]{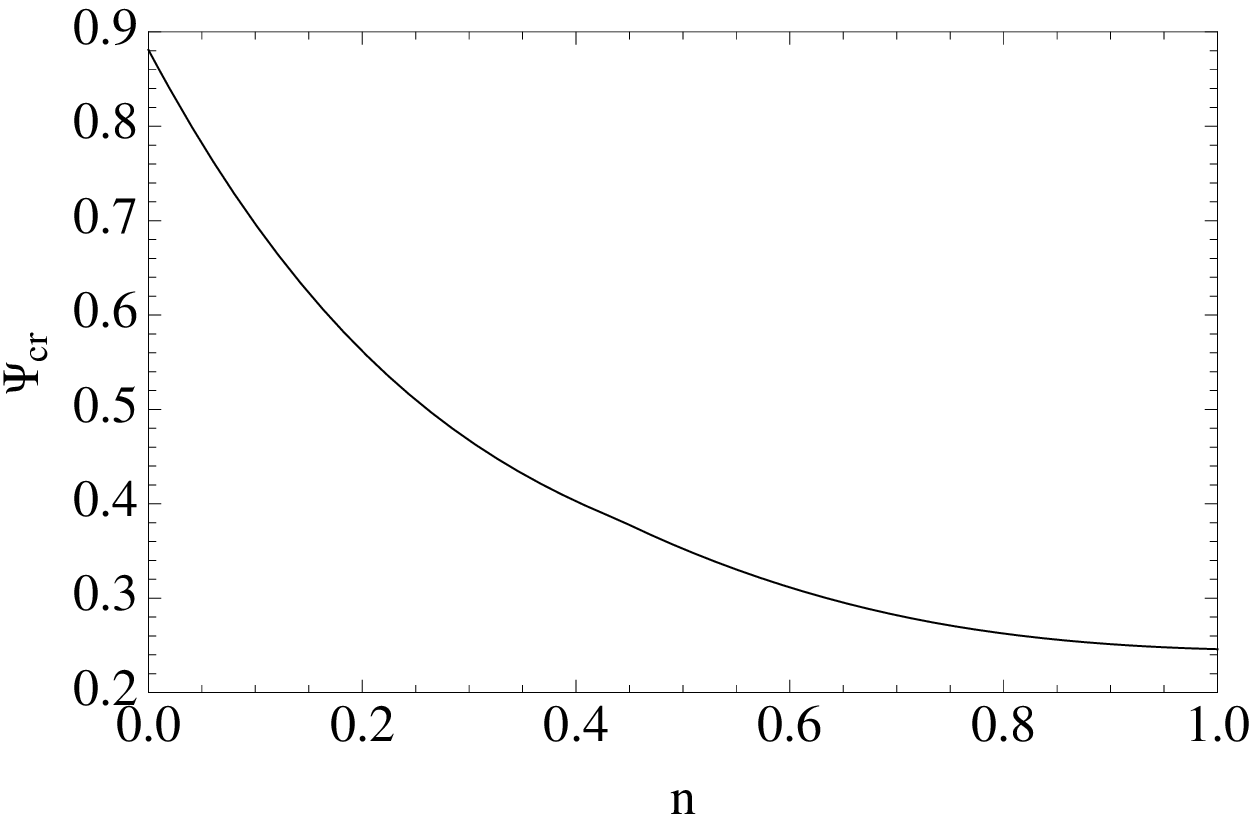}
}
\subfigure[$|u|=2.0$]{
\includegraphics[width=0.75\linewidth]{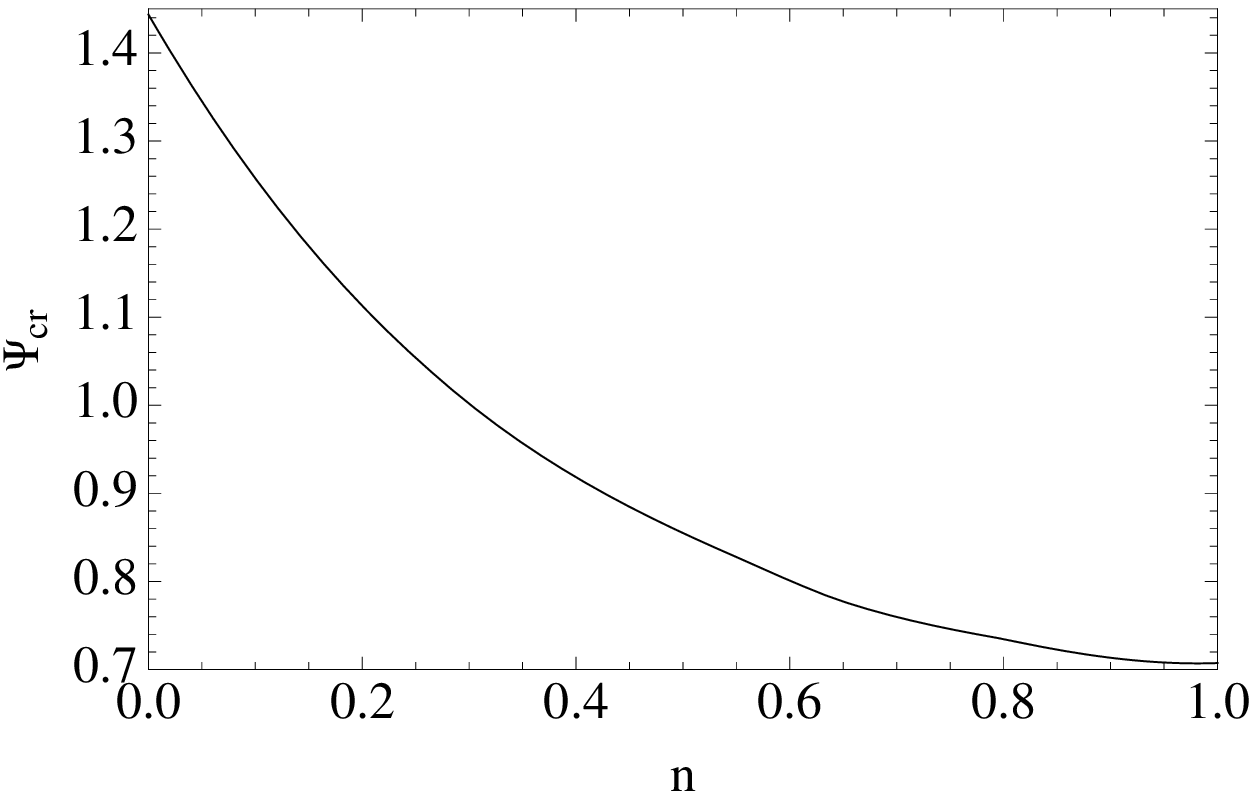}
}
\caption{$\Psi_{\text{cr}}$ in several values of $|u|$
as a function of filling $n$.}
\label{critical-twist}
\end{figure} 
We hereafter consider the bulk limit, namely, $N\to\infty$,
$L\to\infty$, and 
$N/L$ to be a constant, and introduce
the ground-state distribution of $\lambda_{\alpha}$ as 
$\displaystyle\sigma_{0}(\lambda_{\alpha})
=1/L(\lambda_{\alpha+1}-\lambda_{\alpha}).$
By taking derivative of Eq. \eqref{eq:gs}, we have
\beq
&&\frac{1}{\pi}\text{Re}\left[\frac{1}{\sqrt{1-(\lambda +i|u|)^2}}
\right]
=\sigma_{0}(\lambda)\nonumber\\
&&\ \ \ \ \ \ +\frac{1}{\pi}\int_{-B}^{B}
d\lambda'\sigma_0(\lambda')\frac{2|u|}{(2|u|)^2+(\lambda-\lambda')^2}.
\label{eq:gs-2}
\eeq
This equation determines the ground-state distribution.
We note that 
$B$ is related to filling $n$ as follows:
\beq
n=\frac{N}{L}=2\int_{-B}^{B}d\lambda\sigma_0(\lambda).
\label{eq:filling}
\eeq
It is straightforward to check that 
when $B=\infty$, $n=1$, that is,  half filling.
Meanwhile, when $B<\infty$, we obtain $n<1$.
In addition, by performing the particle-hole transformation,
the results of $n>1$ can be obtained from those of $n<1$ \cite{essler2}.
Therefore, in what follows, we only analyze the cases of $n\le 1$.

As for the spin excitation, by introducing the counting functions
$z_c(k_j)=I_j/L$ and $z_s(\lambda_{\alpha})=J_{\alpha}/L$,
Eqs. \eqref{eq:excitation-1} and \eqref{eq:excitation-2} reduce to
\begin{gather}
z_{s}(\lambda)=\frac{1}{\pi}\text{Re}\left[\sin^{-1}(\lambda
 +i|u|)\right]
+\frac{1}{2\pi L}
\sum_{j=1}^{2}\theta(\lambda-\sin k_j)\nonumber\\
+\frac{1}{2\pi}\int_{-B}^{B}d\lambda'
\theta\left(\frac{\lambda-\lambda'}{2}\right)\sigma(\lambda'),
\label{eq:excitation-3}\\
z_{c}(k)=\frac{k}{2\pi}-\frac{i\Psi}{2\pi}+\frac{1}{2\pi}
\int_{-B}^{B}d\lambda\theta(\sin k-\lambda)\sigma(\lambda),
\label{eq:excitation-4}
\end{gather}
where 
$\displaystyle\sigma(\lambda)=\sigma_0(\lambda)
+[\sigma_1^{k_1}(\lambda)
+\sigma_1^{k_2}(\lambda)]/L$ with the shift of the distribution 
$\sigma_1^{k_j}(\lambda)$.
By taking derivative of Eq. \eqref{eq:excitation-3},
we obtain
\beq
\sigma(\lambda)&=&\frac{1}{\pi}\text{Re}\left[
\frac{1}{\sqrt{1-(\lambda +i|u|)^2}}\right]\nonumber\\
&&  +\frac{1}{\pi L}
\sum_{j=1}^{2}\frac{|u|}{|u|^2+(\lambda-\sin k_j)^2}\nonumber\\
&& -\frac{1}{\pi}
\int_{-B}^{B}d\lambda'\frac{2|u|\sigma(\lambda')}
{(2|u|)^2+(\lambda-\lambda')^2}.
\label{eq:density-2}
\eeq
We note that this equation tells us when the analytic properties of the system
change.
In fact, there is a possibility that the last term of the right-hand side
has poles in the complex $k$ plane. 
Since the poles of the nearest to the real axis are 
$\pm i\sinh^{-1}|u|$, the critical value $b_{\text{cr}}$ 
($=\text{Im}(k_{\text{cr}})$) is
\beq
b_{\text{cr}}=\sinh^{-1}|u|,
\eeq
which is equal to the condition that the Mott insulator is broken in the
repulsive Hubbard ring \cite{fukui}.
If $b\equiv\text{Im}(k) $ is less than $b_{\text{cr}}$,
the system is still in the spin-gapped phase 
in which the distribution is obtained from
Eq. \eqref{eq:gs-2}. 
In contrast, at the critical value $b_{\text{cr}}$,
the analytic properties of the system change.
Therefore the breakdown of the spin-gapped phase, namely,
the spin-depairing transition occurs.

A relation between $\Psi$ and $b$ is obtained from
Eq. \eqref{eq:excitation-4}.
By using the fact that the maximum of $\text{Im}(k)$ 
is located on $\text{Re}(k)=0$ and $z_c(k=ib)=0$, 
and neglecting $1/L$ corrections on the distribution,
we have
\beq
\Psi=b-i\int_{-B}^{B}d\lambda\theta(\lambda+i\sinh b)
\sigma_0(\lambda),
\label{eq:critical-psi}
\eeq
which determines the critical value of the twist, $\Psi_{\text{cr}}$. 
We note that the above equation is equal to that appearing 
in the breakdown of the
Mott insulator in the repulsive case \cite{fukui} if $B=\infty$, which
corresponds to the case without imbalance.
Figure \ref{critical-twist} shows $\Psi_{\text{cr}}$ in several values
of $|u|$ as a function
of filling and represents 
that $\Psi_{\text{cr}}$ increase with decreasing filling
in any $|u|$. 
This implies that the spin-gapped state becomes robust
as 
filling is decreased, and is consistent with Fig. \ref{spin-distribution},
which indicates that $b$ decreases with decreasing filling.
In addition, while the increase of $\Psi_{\text{cr}}$ for small $|u|$ 
is tiny at large filling and sharply-rising 
at small filling, that for large $|u|$ is always smooth. 
This is because for large $|u|$, the main contribution of 
Eq. \eqref{eq:critical-psi} comes from the first term
but for small $|u|$ that is not so.

\subsection*{Spin Gap}
\begin{figure}
\subfigure[$|u|=0.25$]{
\includegraphics[width=0.75\linewidth]{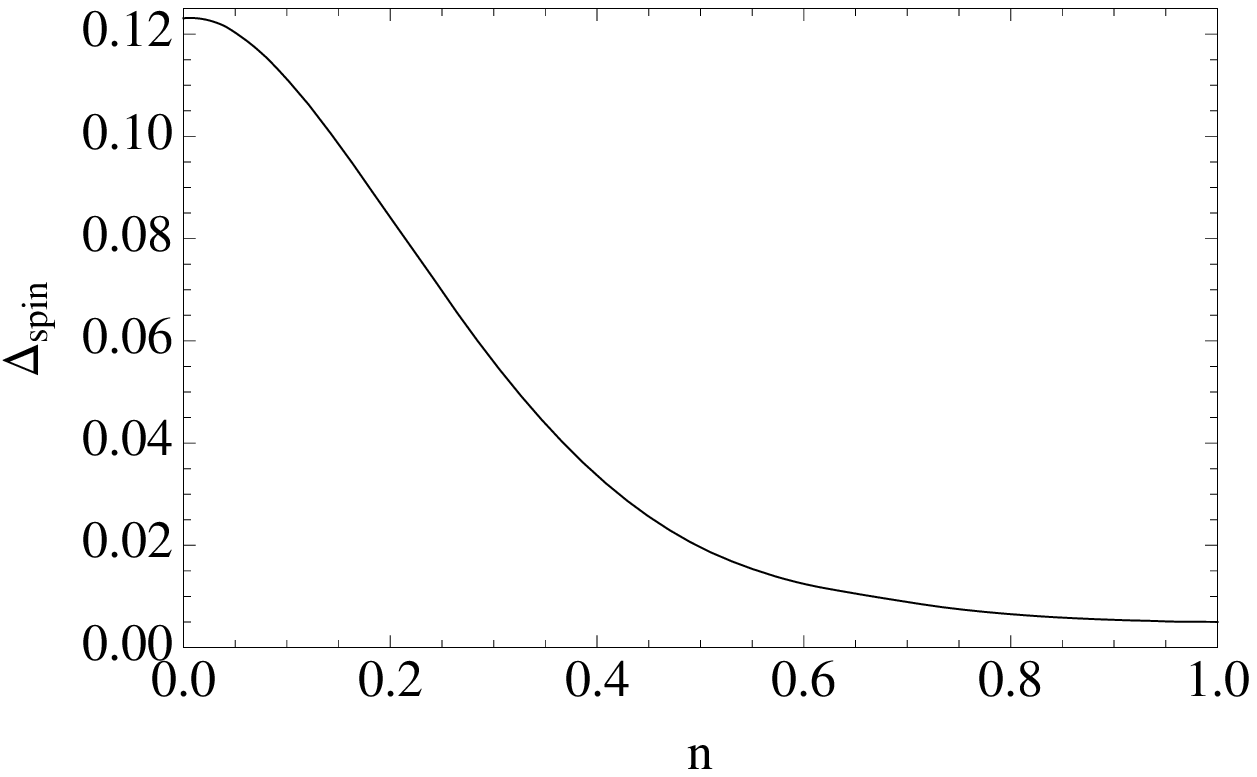}
}
\subfigure[$|u|=0.5$]{
\includegraphics[width=0.75\linewidth]{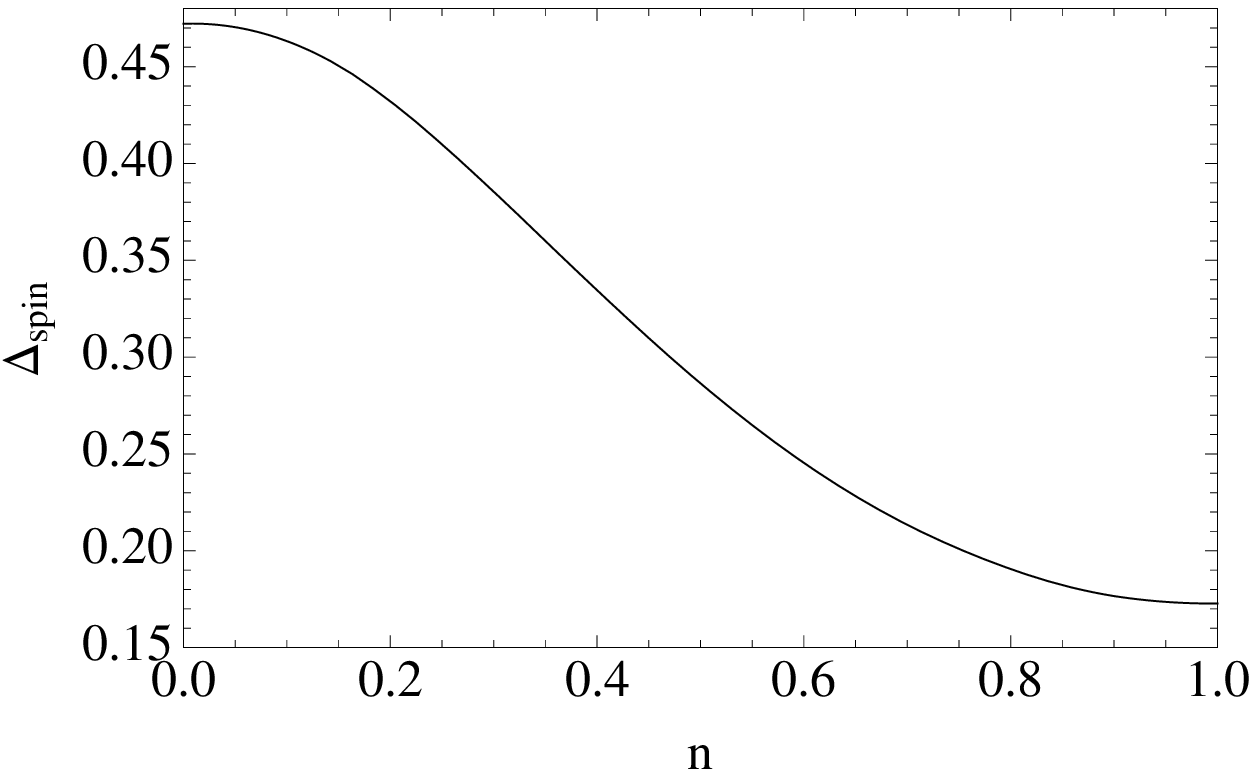}
}
\subfigure[$|u|=1.0$]{
\includegraphics[width=0.75\linewidth]{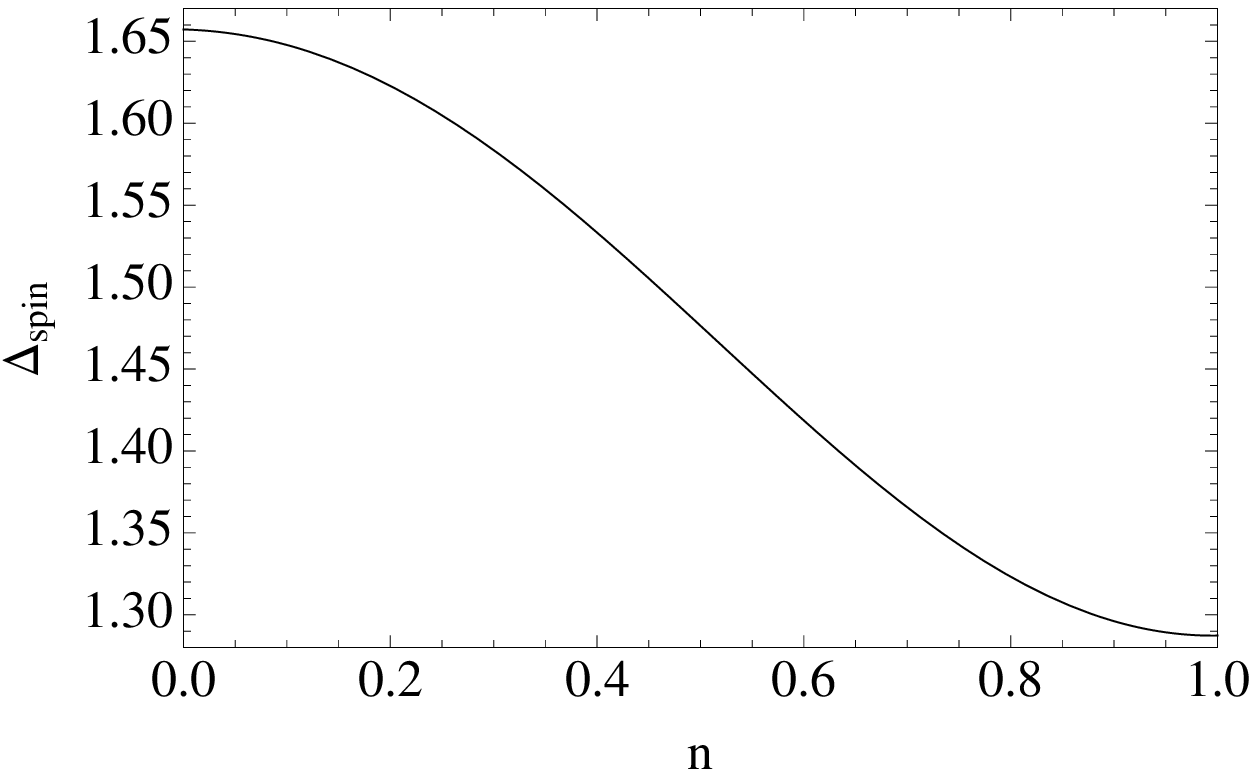}
}
\subfigure[$|u|=2.0$]{
 \includegraphics[width=0.75\linewidth]{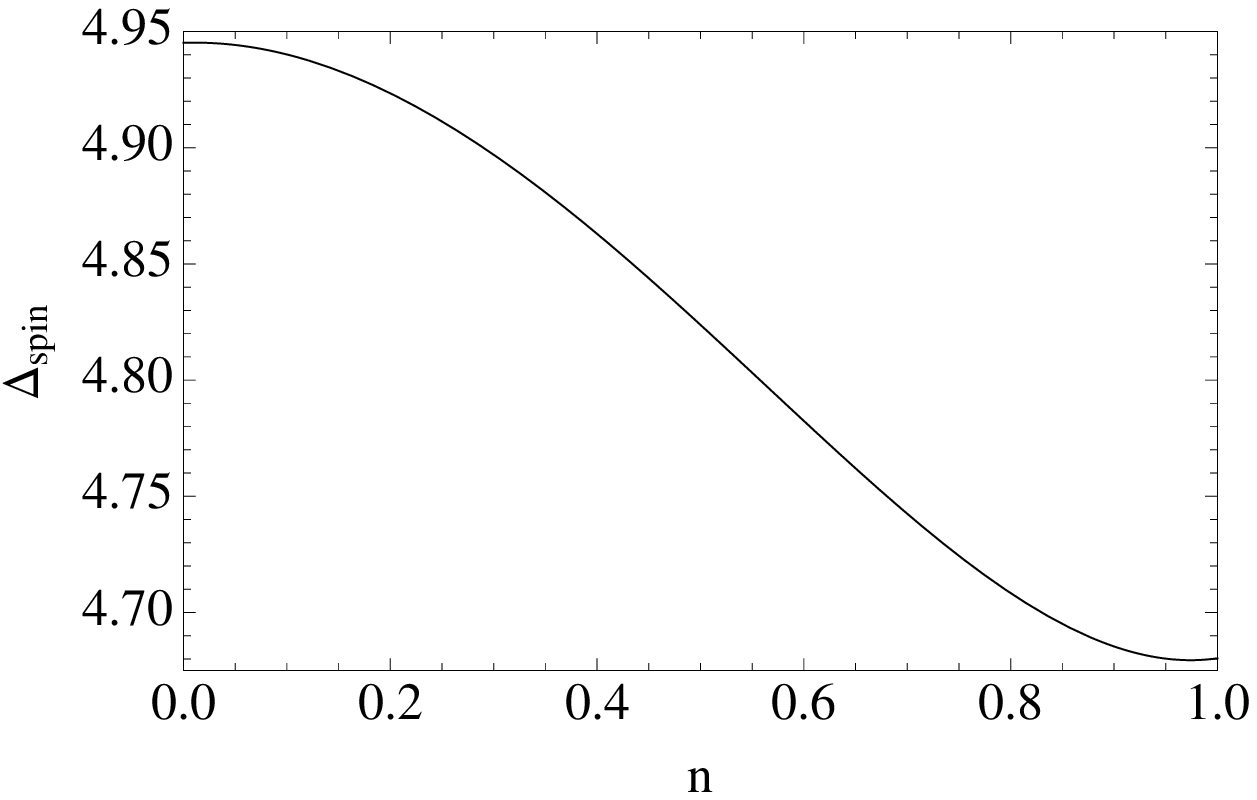}
}
\caption{Spin gap as a function of filling $n$ at $\Psi=0$. }
\label{spin-gap}
\end{figure} 
We now evaluate the spin gap based on the above arguments.
Although the analytical expressions can be obtained at half filling 
with the Fourier transformation
as in Ref. \cite{fukui}, we wish to consider arbitrary
filling cases and continuum model.
To this end, we adopt an approach with the dressed energy
\cite{woynarovich,essler2}, where the elementary excitations 
over the ground state are
obtained by solving the following equations:
\begin{gather}
\epsilon(\lambda)=-4\text{Re}\sqrt{1-(\lambda-i|u|)^2}-2\mu\nonumber\\
\ \ \ \ \ \ \  \ \ \ \ \ \ \ \ \ \  \
-\frac{1}{\pi}\int_{-B}^{B}d\lambda'\frac{2|u|\epsilon(\lambda')}
{(2|u|)^2+(\lambda-\lambda')^2},
\label{eq:dressed-energy-1}\\
e(k)=-2\cos
 k-\mu-\frac{1}{\pi}\int_{-B}^{B}d\lambda\frac{|u|\epsilon(\lambda)}
{|u|^2+(\sin k-\lambda)^2}.
\label{eq:dressed-energy-2}
\end{gather}
Here, $\mu$ is the chemical potential, $-\epsilon(\lambda)$ 
denotes the charge (pair) 
excitation, which satisfies $\epsilon(\pm B)=0$, and 
$e(k)$ denotes the spin (particle) excitation.
We note that when $k=ib_{\text{cr}}$, Eq. \eqref{eq:dressed-energy-2}
is nonanalytic, which is consistent with the argument below Eq. 
\eqref{eq:density-2}. 
This is due to the fact that Eqs. \eqref{eq:dressed-energy-1} and 
\eqref{eq:dressed-energy-2} are related to
Eqs. \eqref{eq:excitation-1}
and \eqref{eq:excitation-2} \cite{woynarovich}.
At half filling, $\mu=-2|u|$ and $B=\infty$, and 
it is straightforward to check that in this case, 
$-\epsilon(\lambda)$ and $e(k)$
correspond to the expressions of the charge and spin
excitations at half filling, respectively.
When $B=0$, the system is in the empty band, and  
the chemical potential is then given by $-2\sqrt{1+|u|^2}$. 
Therefore, possible values of chemical potential are 
\beq
-2\sqrt{1+|u|^2}\le\mu\le -2|u|.
\label{eq:chemical-potential}
\eeq
The spin-triplet excitation is then given by
\beq
\Delta E(k_1,k_2)=e(k_1)+e(k_2)=-2\mu-\sum_{j=1}^2\Bigg[
2\cos k_j \nonumber\\
+\frac{1}{\pi}\int_{-B}^{B}d\lambda
\frac{|u|\epsilon(\lambda)}{|u|^2+(\sin k_j-\lambda)^2}
\Bigg].
\label{eq:spin-excitation-bhf}
\eeq
Figure \ref{spin-gap} depicts the spin gap with $\Psi=0$
in several values of $|u|$ as a function of $n$ and shows that
the spin-gapped state is robust at small $n$ with the fixed $|u|$.
This is because the relative effect of $|u|$ is significant
and it is expected that the tightly-bound spin pairs are 
formed at small $n$.
In the numerical calculation, we first determine $\epsilon(\lambda)$
by solving Eq. \eqref{eq:dressed-energy-1} with the iteration method,
and then calculate the spin gap
by assigning values to Eq. \eqref{eq:dressed-energy-2}. 
By substituting $k_j=ib$ into Eq. \eqref{eq:spin-excitation-bhf}, 
the spin
gap at $\Psi\le\Psi_{\text{cr}}$ is  given by
\beq
\Delta_{\text{spin}}(b)&=&-2\mu-4\cosh b\nonumber\\
&&-\frac{2}{\pi}\int_{-B}^{B}d\lambda
\frac{|u|\epsilon(\lambda)}{|u|^2+(i\sinh b-\lambda)^2}.
\label{eq:spin-gap-bhf}
\eeq

\subsection*{Gap Closing at the Critical Twist}
As discussed above, since 
the analytic properties of the
ground state change at the critical twist,
the spin gap should close at the value.
We analytically show this as follows.
When $b=b_{\text{cr}}-0$, we have
\beq
\frac{|u|}{|u|^2+(i\sinh b-\lambda)^2}
=\frac{1}{2}\Bigg(\text{p.v.}\frac{i}{\lambda}+\pi\delta(\lambda)\nonumber\\
+\frac{1}{2|u|-\eta+i\lambda}\Bigg),
\eeq
where $0<\eta\ll 1$, p.v. denotes the principal value, and we used
$\displaystyle 1/(x+i\eta)=\text{p.v.}(1/x)-i\pi\delta(x).$
By taking into account the property that 
the integral including the principal value
vanishes,
we obtain
\beq
\Delta_{\text{spin}}(b_{\text{cr}}-0)
=-2\mu-4\sqrt{1+(|u|-\eta)^2}-\epsilon(0)\nonumber\\
-\frac{1}{\pi}
\int_{-B}^{B}d\lambda\frac{\epsilon(\lambda)}{2|u|-\eta+i\lambda}.
\label{eq:proof-1}
\eeq
On the other hand, Eq. \eqref{eq:dressed-energy-1} indicates
\beq
\epsilon(0)=-2\mu-4\sqrt{1+|u|^2}-\frac{1}{\pi}\int_{-B}^{B}
d\lambda\frac{\epsilon(\lambda)}{2|u|+i\lambda}.
\label{eq:proof-2}
\eeq
From Eqs. \eqref{eq:proof-1} and \eqref{eq:proof-2},
we conclude that $\Delta_{\text{spin}}(b_{\text{br}})=0$
at arbitrary filling,
which ensures that the spin-depairing transition occurs
at the critical twist.

\section{Transition Probability}
\begin{figure}
\subfigure[$|u|=0.25$]{
\includegraphics[width=0.75\linewidth]{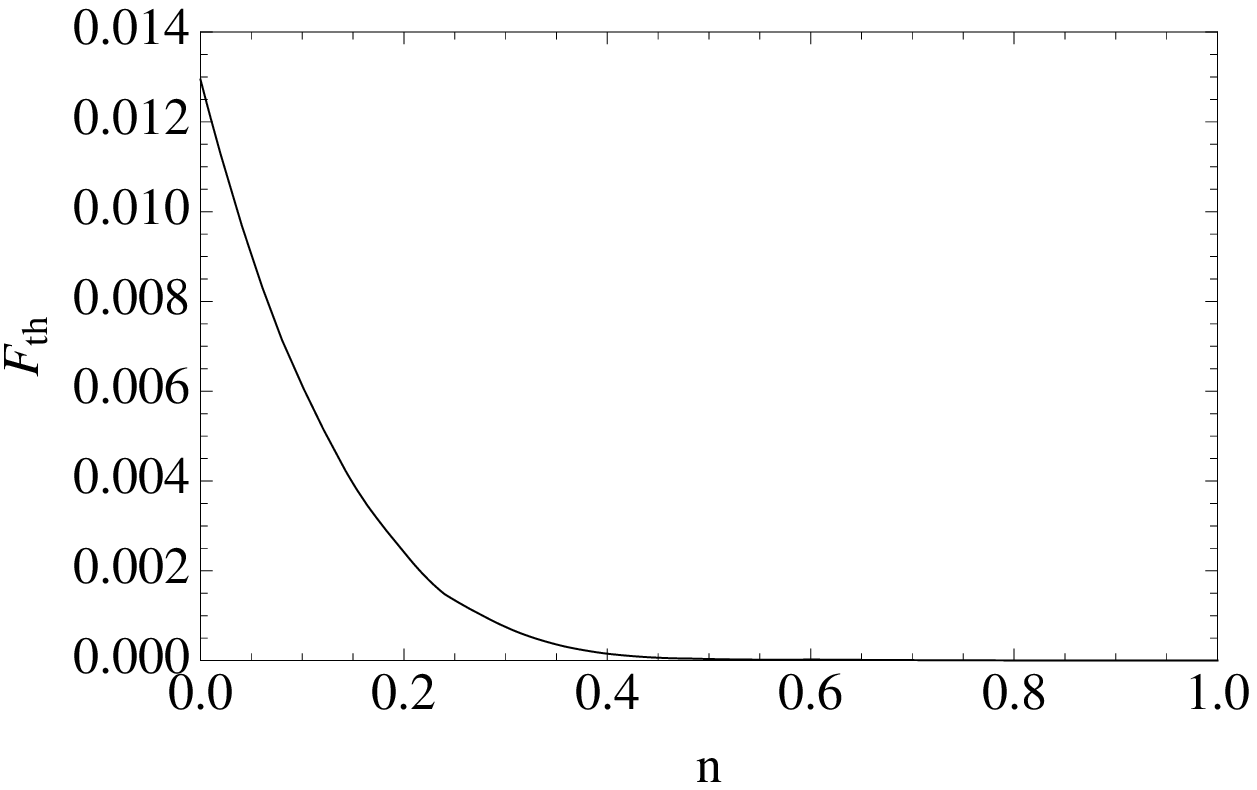}
}
\subfigure[$|u|=0.5$]{
\includegraphics[width=0.75\linewidth]{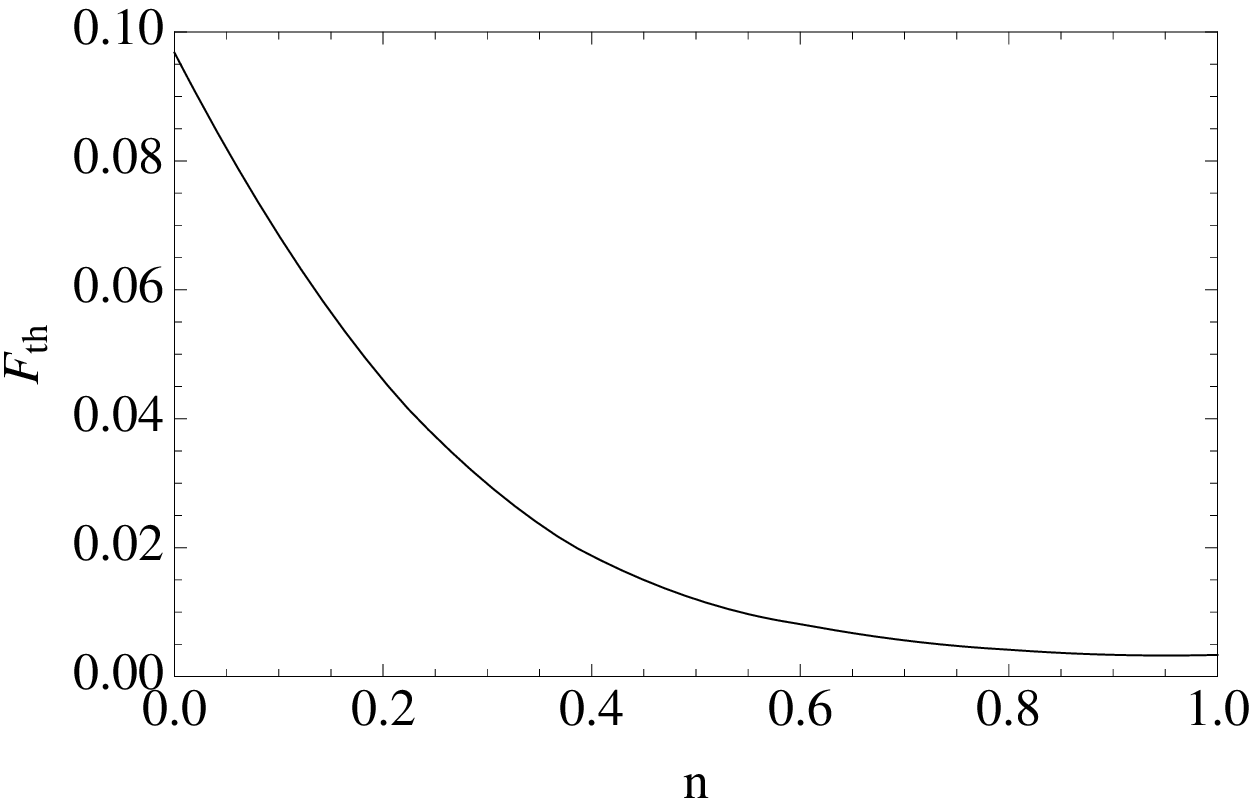}
}
\subfigure[$|u|=1.0$]{
\includegraphics[width=0.75\linewidth]{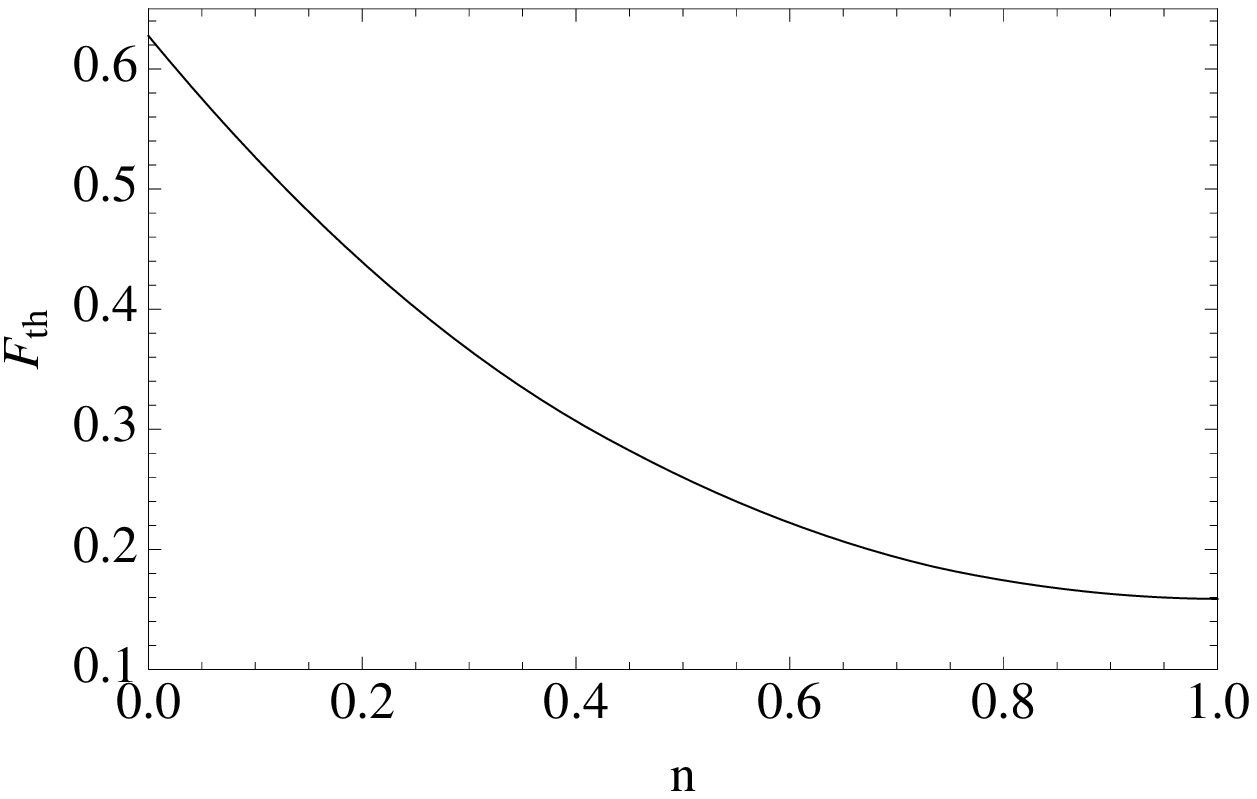}
}
\subfigure[$|u|=2.0$]{
 \includegraphics[width=0.75\linewidth]{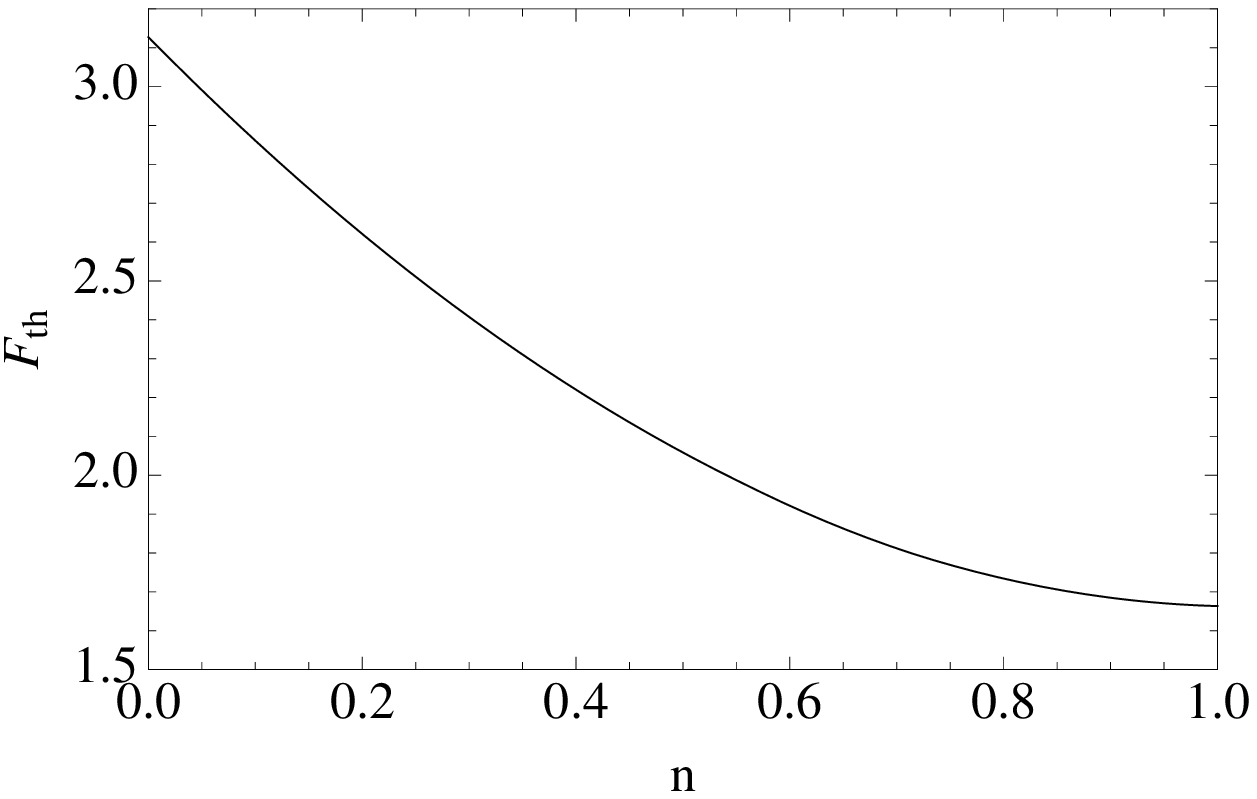}
}
\caption{Value of the threshold field strength. In (a), 
$F_{\text{th}}\ne 0$ in every filling while
it takes a small value. }
\label{transition}
\end{figure}

As first discussed in Ref. \cite{oka},
the Landau-Dykhne method \cite{landau,dykhne,davis} 
makes it possible to relate
the original Hermitian problems \eqref{eq:so-hubbard2} and
\eqref{eq:twist} to
those of the Hubbard ring with complex twists analyzed 
in Sec.  I\hspace{-.1em}I\hspace{-.1em}I.
To this end, let us consider a time-dependent gauge
flux $\Phi(t)=Ft$ with a constant $F$ 
in Eqs. \eqref{eq:so-hubbard2} and \eqref{eq:twist}, which 
is switched on at $t=0$.
Then, based on the analytic continuation of the 
energy levels as a function of complex time,
the Landau-Dykhne method enables us to evaluate
the spin-depairing transition probability, which 
is given by 
\beq
P=\exp{\left[-2\ \text{Im}\int_{0}^{t^*}dt'\Delta_{\text{spin}}
(\Phi(t'))\right]},
\eeq
where $t^{*}$ is the complex time in which the level crossing occurs.
As pointed out in Ref. \cite{oka}, the above integral path lies on
the imaginary axis in the bulk limit. This implies
the relationship $\Phi=i\Psi$ in the same limit, which reduces to
the Hubbard ring with complex twists. 
Therefore, the transition probability becomes 
$\displaystyle P=e^{-\pi F_{\text{th}}/F}$
with
\beq
&&F_{\text{th}}
=\frac{2}{\pi}\int_0^{b_{\text{cr}}}\Delta_{\text{spin}}(b)
\frac{d\Psi}{db}{db}\nonumber\\
&&=\frac{4}{\pi}\int_0^{\sinh^{-1}|u|}
\left[1- \int_{-B}^{B}d\lambda\frac{2|u|\cosh b\sigma_0(\lambda)
}{|u|^2+(i\sinh b-\lambda)^2}\right]\nonumber\\
&&\left[-\mu-2\cosh b- \frac{1}{\pi}\int_{-B}^{B}d\lambda 
\frac{|u|\epsilon(\lambda)}
{|u|^2+(i\sinh b-\lambda)^2}\right]db,
\nonumber\\
\eeq
where 
$F_{\text{th}}$ is the threshold field.
Figure \ref{transition} plots the behavior of $F_{\text{th}}$ as 
a function of filling and shows that the system is vulnerable to
the transition toward  half filling.
In addition to this,
the increase of $F_{\text{th}}$ for 
large $|u|$ is smooth, while that for small $|u|$ has a large curvature
around some filling.
This may be because
while for large $|u|$ the spin pairs are tightly bounded 
at arbitrary filling,
for small $|u|$ the spin pairs are weakly bounded around
half filling but tightly bounded below some filling.

We finally comment on the ground-state decay rate $\Gamma$, which is 
is related to the transition probability  as follows \cite{oka,oka2}:
\beq
\Gamma/L=-\frac{aF}{2\pi}\ln[1-P],
\eeq
with an empirical factor $a$ to be of the order of 1
and to describe the suppression of the tunneling.
In the repulsive Hubbard case
at half filling, this empirical factor 
has been attributed to the pair-annihilation processes
and determined 
as a function of $u$ with the time-dependent 
density matrix renormalization group \cite{oka3}.
On the other hand, in the attractive Hubbard case at 
arbitrary filling, $a$ is still needed to describe the
particle-annihilation processes
but has yet to be determined. The determination of $a$ 
may be done in the same way as  \cite{oka3}.   

\section{Summary and Outlook}
We have investigated the attractive Hubbard ring with
a time-dependent gauge flux on the spin sector 
combining the methods of the Bethe ansatz with
complex twists and Landau-Dykhne, 
and shown that the spin-depairing transition occurs.
We point out that
this transition may 
represent the nonequilibrium transition from fermionic superfluids
to normal states with spin currents
triggered by a many-body quantum tunneling.
By using the formulation with the dressed energies, 
it has been shown that 
as the spin gap increases with decreasing
filling, the threshold field strength
in which the spin-depairing transition occurs 
becomes large as filling is decreased.
Although our analysis has been performed  for
the specific time-dependent gauge flux as $\Phi(t)=Ft$,
taking into account the fact that the breakdown of the Mott insulator 
occurs for another time-dependent gauge flux \cite{oka2},
the same consideration may be obtained in 
the spin-depairing transition. 

We note that the analysis in
the Hubbard ring so far can be directly applied to 
the case of a continuum model,
\beq
H=\int dx\Big[-\sum_{\sigma}\phi^{\dagger}_{\sigma}(x)
\left(\partial_x+i\sigma A(x,t)\right)^2
\phi_{\sigma}(x)\nonumber\\
-4|u|\phi^{\dagger}_{\uparrow}(x)
\phi^{\dagger}_{\downarrow}(x)
\phi_{\downarrow}(x)\phi_{\uparrow}(x)\Big].
\label{eq:continuum-model}
\eeq
This is because effects of gauge fields are again
replaced by boundary conditions \cite{byers}.
The continuum model
is reached by passing to limits $|u|\to 0$ and $n\ll 1$ in
the Hubbard model,
which is due to the fact that mathematically 
the continuum model corresponds to the continuum limit 
of the Hubbard model \cite{essler2}.
By considering the above limits, we have in the continuum model,
$\sinh b\to b$ in Eq. \eqref{eq:critical-psi},
$\text{Re}\sqrt{1-(\lambda-i|u|)^2}\to 1-(\lambda^2+|u|^2)/2$
in Eq. \eqref{eq:dressed-energy-1}, and
$\cos k\to 1-k^2/2$ and $\sin k\to k$ in
Eq. \eqref{eq:dressed-energy-2}.
Therefore, we do not need a special consideration
to calculate the transition probability in the continuum model.

We also point out that although we have analyzed the case of
the spin SO(2) gauge field for experimental simplicity,  
the same treatment is applicable to cases of SU(2) gauge fields since
effects of the gauge fields are again incorporated into boundary conditions 
\cite{fujimoto}. 
As far as the continuum models are concerned, in general,
we obtain exact solutions by the Bethe ansatz method
for attractive SU($N$) Fermi gases on a ring
in which the spin excitations have gaps \cite{schlottmann}.
Hence, the spin-depairing transition in the 
SU($N$) Fermi gases is also expected, which may be realized with,
for example, ytterbium fermionic isotopes \cite{taie}.

\section*{Acknowledgements}
SU is supported by the Grant-in-Aid for the Global COE
Program ``The Next Generation of Physics, Spun from Universality and
Emergence'' from the Ministry of Education, Culture, Sports, Science and
Technology (MEXT) of Japan.
NK is partly supported by Grant-in-Aid for Scientific Research [Grant
nos. 21540359, 20102008] and JSPS through the ``Funding Program for
World-Leading Innovative R\&D on Science and Technology (FIRST
Program)'', initiated by the Council for Science and Technology Policy
(CSTP).

\end{document}